\documentclass[%
reprint, 
superscriptaddress,
nobibnotes,
amsmath,amssymb,
aps,
prl,
floatfix,
]{revtex4-2}

\usepackage{mathtools}
\usepackage{hyperref}
\usepackage{lipsum}
\usepackage{csquotes}
\usepackage{graphicx}
\usepackage{dcolumn}
\usepackage{bm}

\usepackage{printlen} 
\usepackage{setspace}
\usepackage[capitalise]{cleveref}
\usepackage{physics}
\usepackage{xspace}
\usepackage{xcolor}

\newcommand{\suppmat}{Supplemental Material\cs\cite{SuppMat}\xspace}

\newcommand{\supprefdensityfmax}{SM~Fig.~S2a\cs\cite{SuppMat}\xspace}
\newcommand{\supprefclustering}{SM~Fig.~S3\cs\cite{SuppMat}\xspace}

\newcommand{\supprefturnover}{SM~Fig.~S4\cs\cite{SuppMat}\xspace}

\usepackage{color}

\usepackage{multirow}

\newcommand\InPartS{\mbox{InPartS}}

\newcommand*{\kernhat}[2]{#2\kern#1\hat{\phantom{#2}}}

\let\oldchi\chi
\let\chi\undefined
\DeclareRobustCommand{\chi}{{\mathpalette\irchi\relax}}
\newcommand{\irchi}[2]{\raisebox{\depth}{$#1\oldchi$}} 

\newcommand{\cs}{~}

\newcounter{myfigpanel}[figure]
\newcounter{myfigpanelonly}[figure]

\newcommand{\panelletter}[1]{\refstepcounter{myfigpanel}\label{#1}\refstepcounter{myfigpanelonly}\label{onlyletter:#1}\alph{myfigpanel}}
\newcommand{\panel}[1]{(\protect\panelletter{#1})}

\ExplSyntaxOn
\NewDocumentCommand{\panelletters}{m}{
\addtocounter{myfigpanel}{1}\alph{myfigpanel}\addtocounter{myfigpanel}{-1}\textendash%
\clist_map_inline:nn{#1}{%
\refstepcounter{myfigpanel}\label{##1}\refstepcounter{myfigpanelonly}\label{onlyletter:##1}%
}%
\alph{myfigpanel}
}

\ExplSyntaxOff

\crefalias{myfigpanel}{figure}
\crefname{myfigpanelonly}{panel}{panels}

\let\origcaption\caption
\let\caption\undefined
\DeclareRobustCommand{\caption}[1]{\origcaption{\protect\setcounter{myfigpanel}{0}\protect\setcounter{myfigpanelonly}{0}#1}}

\newcommand{\MPIDS}{\affiliation{Max Planck Institute for Dynamics and Self-Organization, Göttingen, Germany}}

\newcommand{\RPCTP}{\affiliation{Rudolf Peierls Centre for Theoretical Physics, University of Oxford, Oxford OX1 3PU, United Kingdom}}
\newcommand{\DIOSCURI}{\affiliation{Dioscuri Centre for Physics and Chemistry of Bacteria, Institute of Physical Chemistry, Warsaw, Poland}}
\newcommand{\UED}{\affiliation{School of Physics and Astronomy, The University of Edinburgh, Edinburgh EH9 3FD, United Kingdom}}

\newcommand{\maxcomp}{\ensuremath{F_{\mathrm{max}}}\xspace}
\newcommand{\motility}{\ensuremath{M}\xspace}
\newcommand{\Dinf}{\ensuremath{D_\infty}\xspace}
\newcommand{\Dt}{\ensuremath{D_{\mathrm{t}}}\xspace}
\newcommand{\vsp}{\ensuremath{v}\xspace}
\newcommand{\trot}{\ensuremath{t_\mathrm{rot}}\xspace}

\begin{document}

\title{Phase separation in a mixture of proliferating and motile active matter}

\author{Lukas Hupe}
\MPIDS
\author{Joanna M. Materska}
\DIOSCURI
\author{David Zwicker}
\MPIDS
\author{Ramin Golestanian}
\email{ramin.golestanian@ds.mpg.de}
\MPIDS
\RPCTP
\author{Bartlomiej Waclaw}
\email{bwaclaw@ed.ac.uk}
\DIOSCURI
\UED
\author{Philip Bittihn}%
\email{philip.bittihn@ds.mpg.de}
\MPIDS

\begin{abstract}
Proliferation and motility are ubiquitous drivers of activity in biological systems.
Here, we study a dense binary mixture of motile and proliferating particles with exclusively repulsive interactions, where homeostasis in the proliferating subpopulation is maintained by pressure-induced removal.
Using computer simulations, we show that phase separation emerges naturally in this system at high density and weak enough self-propulsion.
We show that condensation is caused by interactions between motile particles induced by the growing phase, and recapitulate this behavior in an effective model of only motile particles with attractive interactions.
Our results establish a new type of phase transition and pave a way to reinterpret the physics of dense cellular populations, such as bacterial colonies or tumors, as systems of mixed active matter.
\end{abstract}

\maketitle

Out-of-equilibrium, active matter systems made of self-driven entities such as motile bacteria, synthetic colloids, and flocking animals exhibit a rich array of collective phenomena, including motility-induced phase separation (MIPS), flocking transitions, and defect formation. While significant progress has been made in understanding phase separation within single-component active matter systems\cs\cite{Wensink2012,golestanian1909phoretic,Gompper2020,Marchetti2013,Cates2015,Solon2018,Shankar2018,Juelicher2018,Golestanian2019}, the interplay between distinct types of active matter remains largely unexplored.

Most theoretical and experimental studies have focused on motile active matter, characterized by self-propelled particles that interact through forces and torques, leading to phenomena ranging from emergent swarming behavior to clustering dynamics\cs\cite{Vicsek1995,Gregoire2004,Toner1995,Theurkauff2012,Buttinoni2013}. Systems featuring growing active matter, where particle populations expand through division and growth\cs\cite{hallatschek_proliferating_2023}, have received comparatively less attention. Notable works in this domain include studies on growing bacterial colonies, as well as investigations into the mechanical feedback and pattern formation in tissues and biofilms\cs\cite{Ranft2010,farrell_mechanically_2013, Grant2014,DellArciprete2018,Yaman2019,Zhang2021,Pollack2022,Isensee2022}. These studies underscore the unique physical constraints and emergent behaviors arising from growth-induced stresses and homeostatic regulation.

However, many naturally-occurring systems feature both motile and growing active matter. For example, biofilms often consist of growing bacterial populations interspersed with motile bacteria that swim through the porous matrix\cs\cite{Steinberg2020MotileSubpopulation,Ravel2022swimming}. Similarly, cancerous tissues exhibit a dynamic interplay between proliferating cells and motile cells undergoing epithelial-to-mesenchymal transition (EMT)\cs\cite{thiery_epithelialmesenchymal_2002}. This highlights the need for a systematic understanding of how growth and motility interact and shape collective behaviors. Previous studies have begun to address related questions, such as the clustering of passive particles in an active bath\cs\cite{Stenhammar_activepassive_2015,Wysocki2016,Dolai_activepassive_2018}, segregation of rods with different levels of activity\cs\cite{McCandlish2012}, multiple species interacting non-reciprocally\cs\cite{Saha2020,OuazanReboul_metaboliccycles_2023,Dinelli_nonreciprocity_2023}, and shape-dependent segregation in growing colonies\cs\cite{ghosh_mechanically-driven_2015}. Moreover, recent work on motile and non-motile or slowly-motile cells further emphasizes the relevance of mixed active matter systems\cs\cite{kolb_active_2020,Braat2024}.

From a theoretical perspective, it is intriguing that the properties underlying known transitions in systems of only motile particles could be fundamentally altered by the presence of a growing phase: For example, in flocking, the ratio of rotational diffusion to velocity governs the transition to collective motion in the presence of aligning interactions\cs\cite{Vicsek1995}. For MIPS, density-dependent self-propulsion leads to phase separation\cs\cite{Cates2015}.
The addition of growing particles introduces new timescales and interactions that could modify these properties, leading to new effective interactions between motile particles, potentially introducing new kinds of collective behavior.

In this work, we propose a simple, idealized model that combines motile and growing active matter to investigate how these two types of activity interact and coevolve.
We show that, when the activity of the growing phase is sufficiently strong compared to the activity of the motile phase, the motile phase undergoes a condensation transition that has not been observed in single-component active matter models.
Our findings shed light on the rich interplay between motility and growth, and demonstrate how important it is to consider these interactions in real-world systems such as biofilms, tissues, and other heterogeneous active matter systems.

\begin{figure*}[t]
\includegraphics[width=1.8\columnwidth]{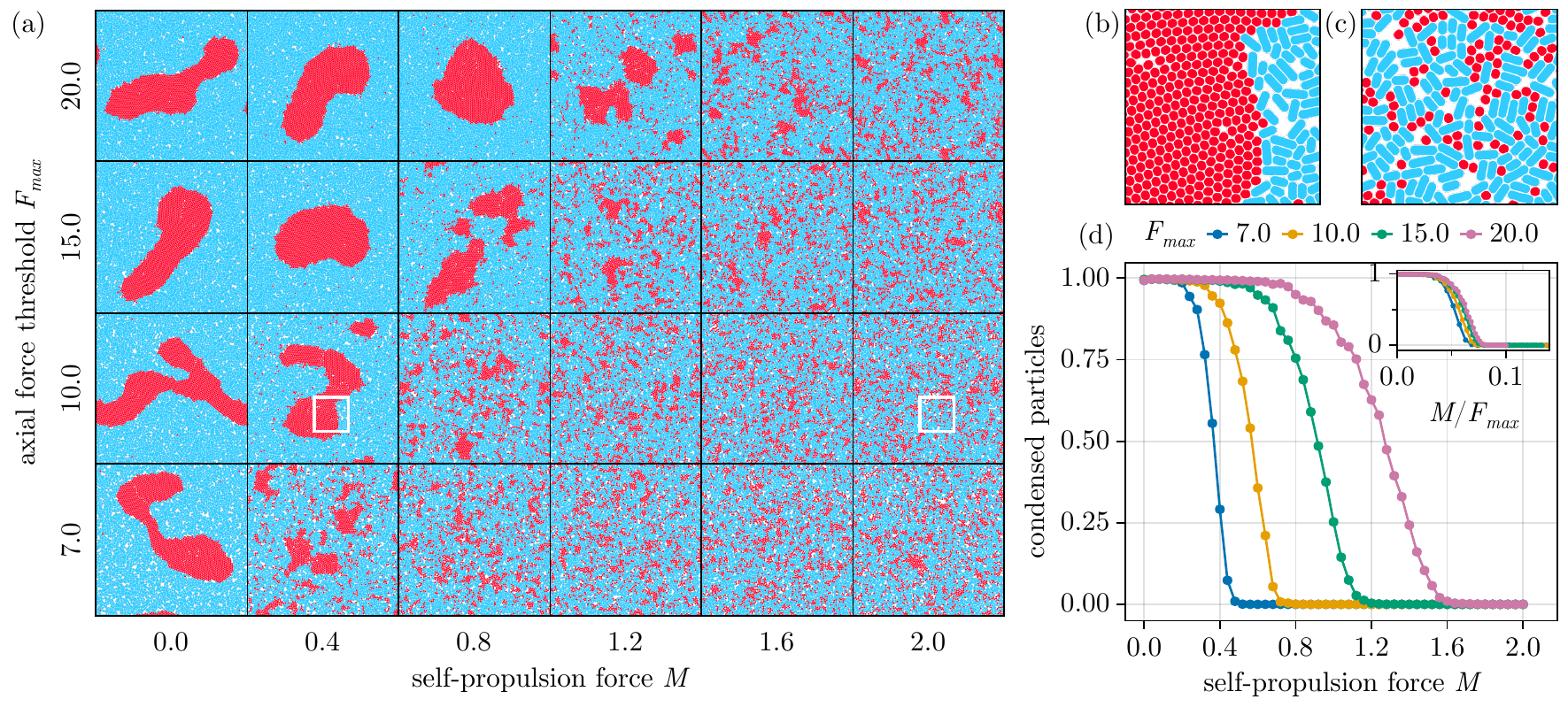}
\caption{
\panel{pan:snapshot_overview}~Snapshots of mixtures of motile (red) and growing (blue) particles at $t=2\times10^4$, for different combinations of self-propulsion force $M$ and axial force threshold $\maxcomp$, showing a transition from a  phase-separated regime (top left) to a mixed regime (bottom right).   White squares represent  regions shown in \cref{onlyletter:pan:zoom_separated,onlyletter:pan:zoom_mixed}.
\panel{pan:zoom_separated}~A zoomed-in view of a phase boundary, for $M = 0.4$ and $\maxcomp = 10.0$.
\panel{pan:zoom_mixed}~A zoomed-in view of the mixed phase, for $M = 2.0$ and $\maxcomp = 10.0$.
\panel{pan:clustersize_selfpropulsion}~Fraction of motile particles in clusters of more than 300 particles as a function of self-propulsion force $M$, for different axial force thresholds $\maxcomp$. Inset: the same data with the self-propulsion force rescaled by $\maxcomp$.
}
\label{fig:fig1}
\end{figure*}

\textit{Model}---We extend the particle-based model~\cite{hupe_minimal_2024} of two-dimensional spherocylinders interacting through steric repulsion and following overdamped dynamics. All particles have the same width, equal to one unit of length.
Growing cells elongate linearly in time, each with a slightly different rate, and divide symmetrically into two cells upon reaching a preset maximum length.
Non-growing, motile cells have a constant, slightly anisotropic shape with an aspect ratio of 1.1 to enable them to rotate in response to external forces.
They self-propel with a force of magnitude $\motility$ along their axis.
Total cell density is kept approximately constant by removing growing cells when their axial compression force exceeds a threshold $\maxcomp$ (see \supprefdensityfmax).
The distribution of the elongation rates is such that the average doubling time is one time unit.
The system is simulated in a square box of size $80\times 80$ with periodic boundaries, for $2\times10^4$ generations. For details, see \suppmat.

\textit{Results.}---Figure~\ref{pan:snapshot_overview} illustrates the final state of these simulations for different values of the self-propulsion $\motility$ and critical compression $\maxcomp$.
We observe that for high compression $\maxcomp$ and low motility $\motility$, motile cells condense into a single large, densely packed cluster (Fig.~\ref{pan:zoom_separated}).
Increasing motility $\motility$
makes the cluster decrease in size and break up, until only a mixed motile-growing phase remains at high $\motility$ (Fig.~\ref{pan:zoom_mixed}).
Measuring the fraction of condensed motile cells quantitatively confirms a transition from full condensation to a mixed state with onset at non-zero $\motility$ (Fig.~\ref{pan:clustersize_selfpropulsion}).
Since $\maxcomp$ modulates the typical forces exerted by growing particles at homeostasis, this indicates that the behavior of motile particles is also modified.
Interestingly, when rescaling the $x$-axis to $\motility/\maxcomp$, the curves collapse (Fig.~\ref{pan:clustersize_selfpropulsion}, inset), suggesting that the transition to phase separation is governed by a balance of motility with the forces exerted by the surrounding growing medium.

Since all explicit interactions in our system rely on contact-based forces, any additional interactions between non-growing cells that could potentially lead to this phase separation must be mediated by the growing medium. To characterize these potential changes induced by non-growing particles, we therefore start with the simplest case: a single non-growing ``tracer'' immersed in the growing medium, which acts as a bath.

Since the removal of particles is triggered by mechanical stress, interactions between the bath and tracer could affect the local turnover rate.
We thus define the local turnover imbalance $\lambda = \kappa_\mathrm{add} - \kappa_\mathrm{div} - \kappa_\mathrm{rem}$ per unit area, where $\kappa_\mathrm{add}$, $\kappa_\mathrm{div}$ and $\kappa_\mathrm{rem}$ are the (position-dependent) rates at which daughter cells are added after division, mother cells are removed for division, and cells are removed due to the compression-induced removal mechanism, respectively~\footnote{To keep the analysis consistent with subsequent velocity calculations, we take the spatial location to be defined by the corresponding particles' center positions}. Fig.~\ref{pan:turnover_local} shows a narrow ring of positive turnover imbalance corresponding to surplus proliferation, surrounded by a wider band of surplus deaths, and finally another band of surplus replications.
These local turnover imbalances are due to divisions in the layer of bath particles immediately in contact with the tracer, as the center positions of dividing and newly born particles must fall within certain distances from the tracer due to their characteristic sizes (small newly born particles vs. large dividing particles).

The cumulative radial turnover imbalance $\Lambda(r) = \frac{1}{\pi r^2}\int_0^r r'\,\mathrm{d}r' \int_0^{2\pi} \lambda(r',\phi)\, \mathrm{d}\phi$ shows that outside the ring structure (beyond approximately $3$ cell widths from the tracer center), turnover remains balanced (Fig~\ref{pan:turnover_cumulative}).
Therefore, turnover imbalances merely correspond to local redistributions of particle positions.
This is confirmed by the radial velocity distribution of bath particles relative to the tracer, which shows movement from regions of surplus additions to surplus removals (Fig.~\ref{pan:bathvelocity}).
Single non-growing particles are therefore able to exert a short-range influence on the bath of growing cells.

\begin{figure}[t]
\includegraphics[width=0.9\columnwidth]{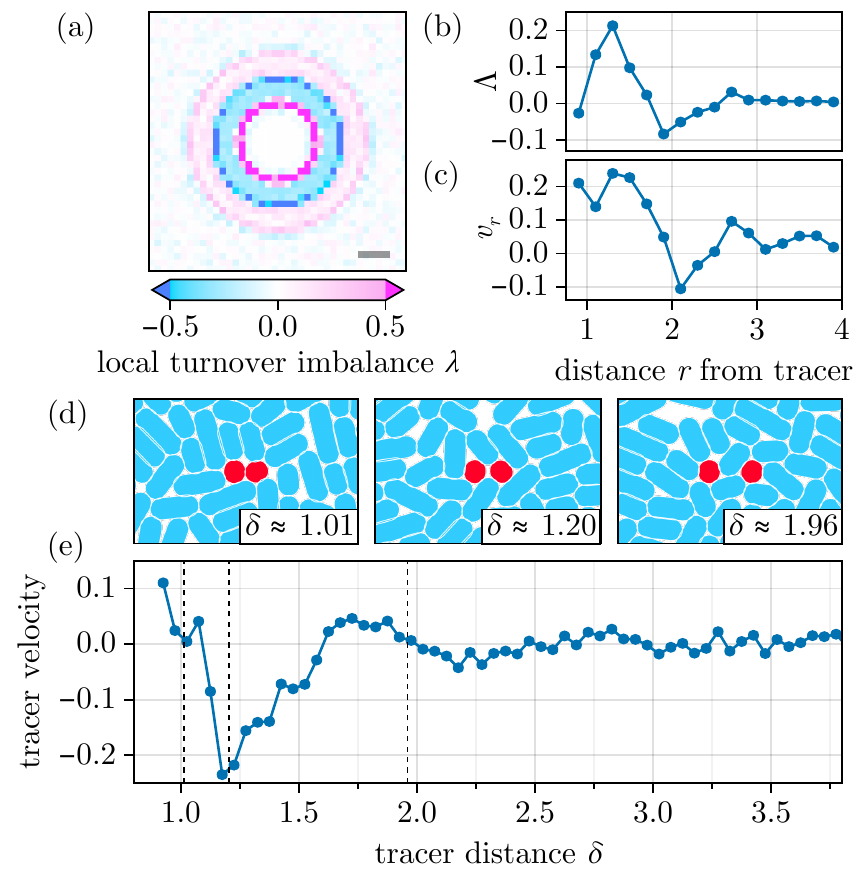}
\caption{
\panel{pan:turnover_local}~Local turnover imbalance $\lambda$ around a single tracer particle. Scale bar shows a distance of one cell width.
\panel{pan:turnover_cumulative}~Integrated turnover imbalance $\Lambda$ as function of distance from tracer center.
\panel{pan:bathvelocity}~Radial bath particle centerposition velocity $v_r$ as a function of distance from tracer center.
\panel{pan:twotracersnaps}~Snapshots of test simulations of two tracers in an active bath for different tracer distances $\delta$.
\panel{pan:twotracerdrift}~Relative tracer drift velocity as a function tracer distance $\delta$.
}
\label{fig:fig2}
\end{figure}

To see if this effect is bi-directional, i.e., perturbations of the growing bath affect other non-growing particles, we introduce a second passive tracer particle into our test simulations.
Measuring the average relative drift velocity of the tracers as a function of tracer center distance $\delta$ (see snapshots in Fig.~\ref{pan:twotracersnaps}), we observe no significant effect beyond $\delta \approx 2$ (Fig~\ref{pan:twotracerdrift}), but a region of negative relative drift for tracer distances between $1$ and $1.6$.

\begin{figure}[t]
\includegraphics[width=0.9\columnwidth]{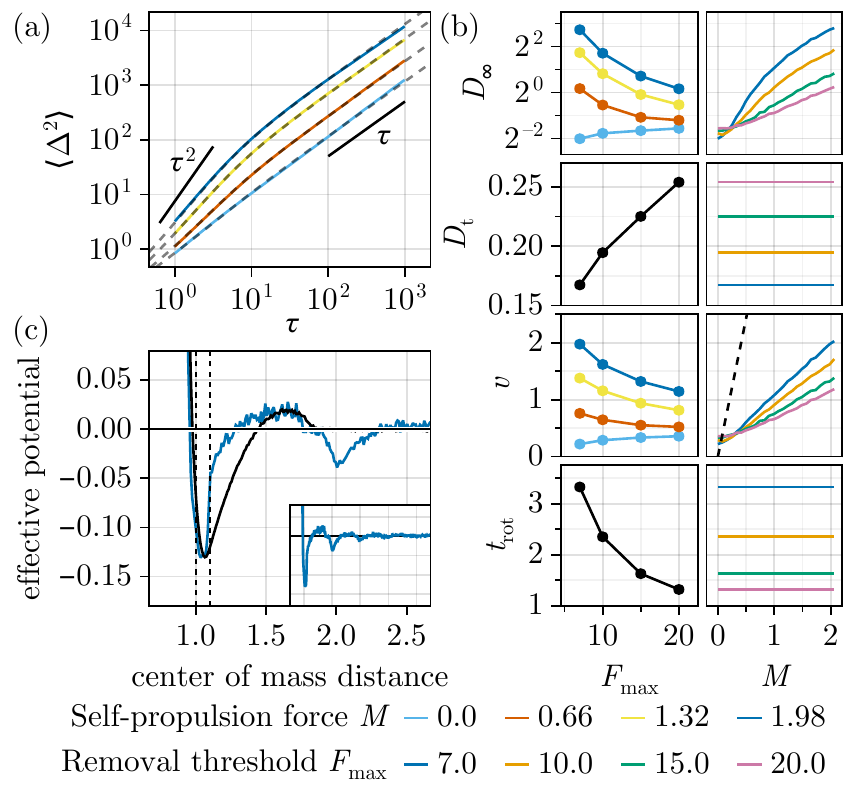}
\caption{
\panel{pan:tracermsd_lowmaxcomp}~Mean squared displacement $\langle\Delta^2\rangle$ of a single self-propelled tracer in a growing bath, computed over $10^5$ generations, for different self-propulsion forces \motility and a fixed $\maxcomp = 10.0$. Dashed lines represent the best-fit active-particle MSD (Eq. (\ref{eq:active})).
\panel{pan:tracermsd_fitparam}~Best-fit parameters (Eq. (\ref{eq:active})):  effective long term diffusion coefficient \Dinf, translational diffusion \Dt, self-propulsion velocity \vsp and persistence time \trot. Fitting was constrained so that \Dt and \trot could not vary with $M$. Dashed black line in the \vsp-plot represents the "bare" velocity of motile particles.
\panel{pan:effectiveinteraction_lowmaxcomp}~The  effective  interaction potential, obtained for two passive tracer particles in a growing bath with $\maxcomp = 10.0$ (blue), and for ABPs using the simple linear adhesion force model (black) with parameters matched to the two-component model. Inset: the potential for longer inter-particle distances upto 5.5.
}
\label{fig:fig3}
\end{figure}

These results reinforce our conjecture that the growing bath introduces effective interactions between individual non-growing particles. In addition, the growing bath particles can also be expected to influence \emph{single} non-growing particles:
First, even passive particles will experience random translational forces and torques due to steric interactions with growing particles. Second, for motile particles, the effective self-propulsion speed will be different from the bare self-propulsion speed of a non-interacting motile particle due to the interaction with their surrounding.

To find out whether these effects of the growing bath are sufficient to enable phase separation and how they conspire quantitatively to do so, we now examine a model with only motile particles, in which interactions with the growing phase are modeled by random, thermal excitations, and a deterministic short-range force captures the effective interaction between pairs of motile particles.
To parameterize this model, we first simulate a single motile particle in the growing bath and obtain its mean squared displacement (MSD) $\langle\Delta^2\rangle(\tau)$ for different $M$ and $\maxcomp$.
Fig.~\ref{pan:tracermsd_lowmaxcomp} shows diffusive motion on long time scales and  ballistic motion at shorter time scales for non-zero motility~$\motility$.
A physically intuitive effective model that can reproduce this behaviour is that of an active Brownian particle (ABP) with translational diffusion $\Dt$, self-propulsion with velocity $\vsp$ and persistence time $\trot$. Its MSD follows~\cite{Howse2007,bechinger_active_2016}
\begin{equation}
\Delta^2(\tau) = \left(4\Dt + 2\vsp^2 \trot\right)\tau + 2\vsp^2 \trot^2 \left(e^{-\tau/\trot} - 1\right), \label{eq:active}
\end{equation}
with a long-term effective diffusion coefficient of $\Dinf = \Dt + \vsp^2\,\trot/2 $.
Fitting this function to the numerically measured mean squared displacement allows us to obtain estimates for the effective ABP parameters as functions of $\maxcomp$ and $\motility$.
To keep the model minimal, we assume that $\Dt$ and $\trot$ are purely determined by interactions with the bath and thus do not vary with $\motility$. Fig.~\ref{pan:tracermsd_lowmaxcomp} shows that this model fits the measured MSD very well.
The inferred parameters are shown in Fig.~\ref{pan:tracermsd_fitparam}.
We observe that the translational diffusion coefficient increases with $\maxcomp$, consistent with higher death-induced activity at higher compression thresholds.
For most values of $\motility$ (except very small ones), the inferred velocity $\vsp$ is lower than the bare self-propulsion velocity (dashed line in the $M$-$v$ plot, Fig.~\ref{pan:tracermsd_fitparam}) due to bath-induced friction. Consistent with this, both the inferred self-propulsion velocity $\vsp$ and persistence time $\trot$ decrease with increasing $\maxcomp$, corresponding to an increased inhibition of persistent self-propulsion at higher densities.
This effect also dominates the effective long-term diffusion coefficient $\Dinf$, which decreases substantially with higher $\maxcomp$ if $\motility$ is not too small.

Interestingly, for very small $\motility$, $\vsp$ does not vanish and the dependence on $\maxcomp$ reverses, so that $\vsp$ increases with density, implying that bath motion is not purely diffusive at short time scales.
This is consistent with prior studies on similar models~\cite{sunkel_motilityinduced_2025, lish_isovolumetric_2024}, which found evidence of ballistic motion at short times and less persistent motion at higher densities.

With the single-particle part of the effective model fully parameterized, we now introduce an effective adhesion force between two ABPs to model the short-range attraction observed in Fig.~\ref{pan:twotracerdrift}.
We neglect the small anisotropy of motile particles and model the adhesion with an isotropic interaction potential $V$ depending only on the distance $\delta$ between particle centers.
Assuming that particles move with mobility $\mu$ and diffusion coefficient $D_\mathrm{t}$ in this potential, it is straight-forward to show using the Fokker-Planck equation that $V(\delta)$ can be obtained as
\begin{equation}
\label{eq:measuredpotential}V(\delta) = -\frac{2D_\mathrm{t}}{\mu}\log{P_\mathrm{s}(\delta)}\,,
\end{equation}
where $P_\mathrm{s}(\delta)$ is the steady-state distribution of inter-particle distances $\delta$ measured in simulations of pairs of tracer particles (averaged over the polar angle according to the assumed rotational symmetry).
\begin{figure}[t]
\includegraphics[width=0.9\columnwidth]{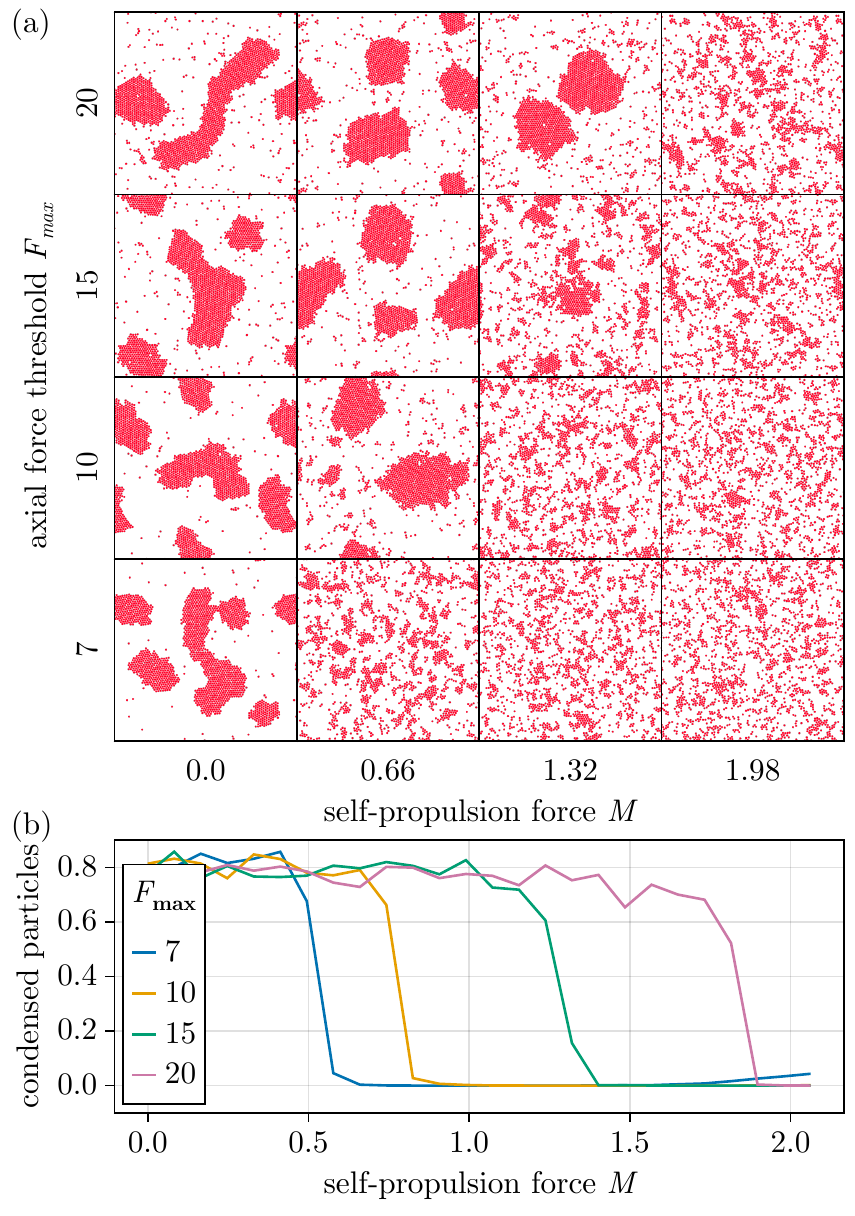}
\caption{
\panel{pan:snapshot_sc}~Snapshots of simulations of the single component model after $2\times10^4$ time units, for parameters corresponding to different combinations of self-propulsion force $M$ and axial force threshold $\maxcomp$, showing a transition from a fully phase-separated regime in the top left to a mixed regime in the bottom right.
\panel{pan:condensed_singlecomp}~Fraction of particles within clusters of at least 300 particles for parameters corresponding to different self-propulsion forces and removal thresholds.
}
\label{fig:fig4}
\end{figure}
Fig.~\ref{pan:effectiveinteraction_lowmaxcomp} shows the effective potential computed in this way for $\maxcomp=10$.
We observe strong attraction near the tracer, followed by a repeating pattern of attraction and repulsion, with a period of approximately $1$ cell width and quickly decaying amplitude, which is likely caused by the finite size of the bath particles. 
However, for simplicity, we shall only model the initial potential well and barrier with a simple linear force law and corresponding potential,
\begin{align}
\label{eq:imposedpotential}V(r) &= -\Delta V\left(\frac{r}{r_2}-1\right)^2 + V_2,
\end{align}
for $0 < r \le r_2\,(1 + \sqrt{V_2/\Delta V})$
with $\Delta V$ and $V_2$ fitted to $V(r)$ from the two-component simulation (see \suppmat for details).
Fig.~\ref{pan:effectiveinteraction_lowmaxcomp} shows excellent agreement between the effective potentials determined via Eq.~\eqref{eq:imposedpotential} in two-component simulations and ABP simulations.

The complete single-component model now consists of the ABP core component incorporating translational diffusion, self-propulsion velocity and persistence time, the linear adhesion model, as well as steric interactions and equations of motion of the existing rod model.
Fig.~\ref{pan:snapshot_sc} shows snapshots of simulations using inferred parameters for different $\motility$ and $\maxcomp$.
We see a clear transition from the condensed to the mixed state, showing that the single component model qualitatively captures the behavior of the full model from Fig.~\ref{pan:snapshot_overview}.
The fact that the particles do not condense into a single cluster within the simulation time is likely due to slow condensation dynamics, because single clusters do not evaporate once formed (\supprefclustering).
The fraction of condensed particles (Fig.~\ref{pan:condensed_singlecomp}) confirms the presence of the phase transition, albeit at slightly different values of $M$ than in the two-component model.

\textit{Discussion}---We have demonstrated that a heterogeneous active mixture of growing and motile particles spontaneously separates into two phases for a range of model parameters. The two types of activity, growth and motility, oppose each other: growing particles exert an effective pressure that tries to condense motile particles, whereas motile particles try to overcome this dynamic confinement.
The onset of condensation is governed by a balance between self‐propulsion and the typical forces exerted by the surrounding growing bath.  Note that this is in some sense the opposite of MIPS, where higher self-propulsion velocities enable condensation, whereas here, self-propulsion \emph{overcomes} an effective attraction that is always present. We also show that a model with only motile particles and an effective attraction qualitatively reproduces the condensation phenomenon.
It will be important to uncover the mechanism giving rise to this attraction. Among other scenarios, an intriguing possibility is to interpret it as a non-equilibrium fluctuation-induced (Casimir-like) force \cite{Casimir1999}, given that all interactions are mediated by random mechanical forces originating from the growing medium.

We also observe quantitative differences between the full and the single-component APB model, most prominently the speed of coarsening after the first clusters have formed. One reason could be that our ABP model only encodes pairwise interactions and thermal noise whose magnitudes are independent of the particles’ environment. In the full model, larger clusters of motile cells might perturb the growing bath more strongly due to their reduced mobility. Indeed, larger non-growing particles mimicking large clusters cause much longer ranged perturbations of the growing bath and advection towards them (see \supprefturnover) that would enhance accretion of additional particles for existing clusters. In the APB model, on the other hand, effective adhesion has always the same (short) range. In addition, in the full model, motile particles within a cluster become shielded from the bath, reducing the effective noise they are exposed to. Also, the true force field fluctuates strongly and is non‐Brownian at short times. Both features are not accounted for in our approach. Exploring these effects for larger clusters might therefore be worthwhile to delineate the contributions of Bose-Einstein-like condensation due to diffusivity-edge~\cite{Golestanian2019,Mahault2020} and other higher-order effects, and make connections with other non-equilibrium models that exhibit condensation \cite{waclaw_mass_2009,waclaw_explosive_2012}.

Finally, many assumptions made here could be relaxed or altered, potentially leading to additional phenomena or substantially changed behavior. For example, we focused on low fractions of motile particles. Exploring the opposite limit, which might be more realistic for, e.g., bacterial sludge bioreactors, may reveal distinct segregation patterns or banding phenomena. We also considered nearly circular motile particles; introducing elongation could lead to nematically ordered clusters. Finally, the homeostatic removal mechanism will depend on biological context and could be generalized to density‐dependent death rates or replaced by entirely random removal, potentially altering fluctuation spectra and effective interactions.
\newline

\acknowledgments
\textit{Acknowledgments}---We acknowledge support from the Max Planck School Matter to Life and the MaxSynBio Consortium, which are jointly funded by the Federal Ministry of Education and Research (BMBF) of Germany and the Max Planck Society.
BW and JM thank Dioscuri, a programme initiated by the Max Planck Society, jointly managed with the National Science Centre in Poland, and mutually funded by Polish Ministry of Science and Higher Education and German Federal Ministry of Education and Research, grant no. UMO-2019/02/H/NZ6/00003.

\textit{Code availability}---Simulations used for this study were built in Julia\cs\cite{bezanson_julia_2017}, using the open-source simulation framework \href{https://www.inparts.org}{\InPartS}\cs\cite{hupe_inparts_2022}. The implementation of the agent-based models will be made available as part of the \InPartS{}Biome model library, published at \url{http://hdl.handle.net/21.11101/0000-0007-FE13-6}.
Example code for simulation and data analysis will be made available alongside the final publication.
Visualisations were prepared using Makie.jl~\cite{danisch_makie_2021}.

\interlinepenalty=10000


\begin{thebibliography}{50}%
\makeatletter
\providecommand \@ifxundefined [1]{%
 \@ifx{#1\undefined}
}%
\providecommand \@ifnum [1]{%
 \ifnum #1\expandafter \@firstoftwo
 \else \expandafter \@secondoftwo
 \fi
}%
\providecommand \@ifx [1]{%
 \ifx #1\expandafter \@firstoftwo
 \else \expandafter \@secondoftwo
 \fi
}%
\providecommand \natexlab [1]{#1}%
\providecommand \enquote  [1]{``#1''}%
\providecommand \bibnamefont  [1]{#1}%
\providecommand \bibfnamefont [1]{#1}%
\providecommand \citenamefont [1]{#1}%
\providecommand \href@noop [0]{\@secondoftwo}%
\providecommand \href [0]{\begingroup \@sanitize@url \@href}%
\providecommand \@href[1]{\@@startlink{#1}\@@href}%
\providecommand \@@href[1]{\endgroup#1\@@endlink}%
\providecommand \@sanitize@url [0]{\catcode `\\12\catcode `\$12\catcode
  `\&12\catcode `\#12\catcode `\^12\catcode `\_12\catcode `\%12\relax}%
\providecommand \@@startlink[1]{}%
\providecommand \@@endlink[0]{}%
\providecommand \url  [0]{\begingroup\@sanitize@url \@url }%
\providecommand \@url [1]{\endgroup\@href {#1}{\urlprefix }}%
\providecommand \urlprefix  [0]{URL }%
\providecommand \Eprint [0]{\href }%
\providecommand \doibase [0]{https://doi.org/}%
\providecommand \selectlanguage [0]{\@gobble}%
\providecommand \bibinfo  [0]{\@secondoftwo}%
\providecommand \bibfield  [0]{\@secondoftwo}%
\providecommand \translation [1]{[#1]}%
\providecommand \BibitemOpen [0]{}%
\providecommand \bibitemStop [0]{}%
\providecommand \bibitemNoStop [0]{.\EOS\space}%
\providecommand \EOS [0]{\spacefactor3000\relax}%
\providecommand \BibitemShut  [1]{\csname bibitem#1\endcsname}%
\let\auto@bib@innerbib\@empty
\bibitem [{\citenamefont {Wensink}\ \emph {et~al.}(2012)\citenamefont
  {Wensink}, \citenamefont {Dunkel}, \citenamefont {Heidenreich}, \citenamefont
  {Drescher}, \citenamefont {Goldstein}, \citenamefont {L\"{o}wen},\ and\
  \citenamefont {Yeomans}}]{Wensink2012}%
  \BibitemOpen
  \bibfield  {author} {\bibinfo {author} {\bibfnamefont {H.~H.}\ \bibnamefont
  {Wensink}}, \bibinfo {author} {\bibfnamefont {J.}~\bibnamefont {Dunkel}},
  \bibinfo {author} {\bibfnamefont {S.}~\bibnamefont {Heidenreich}}, \bibinfo
  {author} {\bibfnamefont {K.}~\bibnamefont {Drescher}}, \bibinfo {author}
  {\bibfnamefont {R.~E.}\ \bibnamefont {Goldstein}}, \bibinfo {author}
  {\bibfnamefont {H.}~\bibnamefont {L\"{o}wen}},\ and\ \bibinfo {author}
  {\bibfnamefont {J.~M.}\ \bibnamefont {Yeomans}},\ }\href
  {https://doi.org/10.1073/pnas.1202032109} {\bibfield  {journal} {\bibinfo
  {journal} {Proceedings of the National Academy of Sciences}\ }\textbf
  {\bibinfo {volume} {109}},\ \bibinfo {pages} {14308–14313} (\bibinfo {year}
  {2012})}\BibitemShut {NoStop}%
\bibitem [{\citenamefont {Golestanian}(2022)}]{golestanian1909phoretic}%
  \BibitemOpen
  \bibfield  {author} {\bibinfo {author} {\bibfnamefont {R.}~\bibnamefont
  {Golestanian}},\ }in\ \href@noop {} {\emph {\bibinfo {booktitle} {{Active
  Matter and Nonequilibrium Statistical Physics: Lecture Notes of the Les
  Houches Summer School: Volume 112, September 2018}}}}\ (\bibinfo  {publisher}
  {Oxford University Press},\ \bibinfo {year} {2022})\BibitemShut {NoStop}%
\bibitem [{\citenamefont {Gompper}\ \emph {et~al.}(2020)\citenamefont
  {Gompper}, \citenamefont {Winkler}, \citenamefont {Speck}, \citenamefont
  {Solon}, \citenamefont {Nardini}, \citenamefont {Peruani}, \citenamefont
  {L\"{o}wen}, \citenamefont {Golestanian}, \citenamefont {Kaupp},
  \citenamefont {Alvarez}, \citenamefont {Kiørboe}, \citenamefont {Lauga},
  \citenamefont {Poon}, \citenamefont {DeSimone}, \citenamefont
  {Muiños-Landin}, \citenamefont {Fischer}, \citenamefont {S\"{o}ker},
  \citenamefont {Cichos}, \citenamefont {Kapral}, \citenamefont {Gaspard},
  \citenamefont {Ripoll}, \citenamefont {Sagues}, \citenamefont
  {Doostmohammadi}, \citenamefont {Yeomans}, \citenamefont {Aranson},
  \citenamefont {Bechinger}, \citenamefont {Stark}, \citenamefont {Hemelrijk},
  \citenamefont {Nedelec}, \citenamefont {Sarkar}, \citenamefont {Aryaksama},
  \citenamefont {Lacroix}, \citenamefont {Duclos}, \citenamefont {Yashunsky},
  \citenamefont {Silberzan}, \citenamefont {Arroyo},\ and\ \citenamefont
  {Kale}}]{Gompper2020}%
  \BibitemOpen
  \bibfield  {author} {\bibinfo {author} {\bibfnamefont {G.}~\bibnamefont
  {Gompper}}, \bibinfo {author} {\bibfnamefont {R.~G.}\ \bibnamefont
  {Winkler}}, \bibinfo {author} {\bibfnamefont {T.}~\bibnamefont {Speck}},
  \bibinfo {author} {\bibfnamefont {A.}~\bibnamefont {Solon}}, \bibinfo
  {author} {\bibfnamefont {C.}~\bibnamefont {Nardini}}, \bibinfo {author}
  {\bibfnamefont {F.}~\bibnamefont {Peruani}}, \bibinfo {author} {\bibfnamefont
  {H.}~\bibnamefont {L\"{o}wen}}, \bibinfo {author} {\bibfnamefont
  {R.}~\bibnamefont {Golestanian}}, \bibinfo {author} {\bibfnamefont {U.~B.}\
  \bibnamefont {Kaupp}}, \bibinfo {author} {\bibfnamefont {L.}~\bibnamefont
  {Alvarez}}, \bibinfo {author} {\bibfnamefont {T.}~\bibnamefont {Kiørboe}},
  \bibinfo {author} {\bibfnamefont {E.}~\bibnamefont {Lauga}}, \bibinfo
  {author} {\bibfnamefont {W.~C.~K.}\ \bibnamefont {Poon}}, \bibinfo {author}
  {\bibfnamefont {A.}~\bibnamefont {DeSimone}}, \bibinfo {author}
  {\bibfnamefont {S.}~\bibnamefont {Muiños-Landin}}, \bibinfo {author}
  {\bibfnamefont {A.}~\bibnamefont {Fischer}}, \bibinfo {author} {\bibfnamefont
  {N.~A.}\ \bibnamefont {S\"{o}ker}}, \bibinfo {author} {\bibfnamefont
  {F.}~\bibnamefont {Cichos}}, \bibinfo {author} {\bibfnamefont
  {R.}~\bibnamefont {Kapral}}, \bibinfo {author} {\bibfnamefont
  {P.}~\bibnamefont {Gaspard}}, \bibinfo {author} {\bibfnamefont
  {M.}~\bibnamefont {Ripoll}}, \bibinfo {author} {\bibfnamefont
  {F.}~\bibnamefont {Sagues}}, \bibinfo {author} {\bibfnamefont
  {A.}~\bibnamefont {Doostmohammadi}}, \bibinfo {author} {\bibfnamefont
  {J.~M.}\ \bibnamefont {Yeomans}}, \bibinfo {author} {\bibfnamefont {I.~S.}\
  \bibnamefont {Aranson}}, \bibinfo {author} {\bibfnamefont {C.}~\bibnamefont
  {Bechinger}}, \bibinfo {author} {\bibfnamefont {H.}~\bibnamefont {Stark}},
  \bibinfo {author} {\bibfnamefont {C.~K.}\ \bibnamefont {Hemelrijk}}, \bibinfo
  {author} {\bibfnamefont {F.~J.}\ \bibnamefont {Nedelec}}, \bibinfo {author}
  {\bibfnamefont {T.}~\bibnamefont {Sarkar}}, \bibinfo {author} {\bibfnamefont
  {T.}~\bibnamefont {Aryaksama}}, \bibinfo {author} {\bibfnamefont
  {M.}~\bibnamefont {Lacroix}}, \bibinfo {author} {\bibfnamefont
  {G.}~\bibnamefont {Duclos}}, \bibinfo {author} {\bibfnamefont
  {V.}~\bibnamefont {Yashunsky}}, \bibinfo {author} {\bibfnamefont
  {P.}~\bibnamefont {Silberzan}}, \bibinfo {author} {\bibfnamefont
  {M.}~\bibnamefont {Arroyo}},\ and\ \bibinfo {author} {\bibfnamefont
  {S.}~\bibnamefont {Kale}},\ }\href {https://doi.org/10.1088/1361-648x/ab6348}
  {\bibfield  {journal} {\bibinfo  {journal} {Journal of Physics: Condensed
  Matter}\ }\textbf {\bibinfo {volume} {32}},\ \bibinfo {pages} {193001}
  (\bibinfo {year} {2020})}\BibitemShut {NoStop}%
\bibitem [{\citenamefont {Marchetti}\ \emph {et~al.}(2013)\citenamefont
  {Marchetti}, \citenamefont {Joanny}, \citenamefont {Ramaswamy}, \citenamefont
  {Liverpool}, \citenamefont {Prost}, \citenamefont {Rao},\ and\ \citenamefont
  {Simha}}]{Marchetti2013}%
  \BibitemOpen
  \bibfield  {author} {\bibinfo {author} {\bibfnamefont {M.~C.}\ \bibnamefont
  {Marchetti}}, \bibinfo {author} {\bibfnamefont {J.~F.}\ \bibnamefont
  {Joanny}}, \bibinfo {author} {\bibfnamefont {S.}~\bibnamefont {Ramaswamy}},
  \bibinfo {author} {\bibfnamefont {T.~B.}\ \bibnamefont {Liverpool}}, \bibinfo
  {author} {\bibfnamefont {J.}~\bibnamefont {Prost}}, \bibinfo {author}
  {\bibfnamefont {M.}~\bibnamefont {Rao}},\ and\ \bibinfo {author}
  {\bibfnamefont {R.~A.}\ \bibnamefont {Simha}},\ }\href
  {https://doi.org/10.1103/revmodphys.85.1143} {\bibfield  {journal} {\bibinfo
  {journal} {Reviews of Modern Physics}\ }\textbf {\bibinfo {volume} {85}},\
  \bibinfo {pages} {1143–1189} (\bibinfo {year} {2013})}\BibitemShut
  {NoStop}%
\bibitem [{\citenamefont {Cates}\ and\ \citenamefont
  {Tailleur}(2015)}]{Cates2015}%
  \BibitemOpen
  \bibfield  {author} {\bibinfo {author} {\bibfnamefont {M.~E.}\ \bibnamefont
  {Cates}}\ and\ \bibinfo {author} {\bibfnamefont {J.}~\bibnamefont
  {Tailleur}},\ }\href
  {https://doi.org/10.1146/annurev-conmatphys-031214-014710} {\bibfield
  {journal} {\bibinfo  {journal} {Annual Review of Condensed Matter Physics}\
  }\textbf {\bibinfo {volume} {6}},\ \bibinfo {pages} {219–244} (\bibinfo
  {year} {2015})}\BibitemShut {NoStop}%
\bibitem [{\citenamefont {Solon}\ \emph {et~al.}(2018)\citenamefont {Solon},
  \citenamefont {Stenhammar}, \citenamefont {Cates}, \citenamefont {Kafri},\
  and\ \citenamefont {Tailleur}}]{Solon2018}%
  \BibitemOpen
  \bibfield  {author} {\bibinfo {author} {\bibfnamefont {A.~P.}\ \bibnamefont
  {Solon}}, \bibinfo {author} {\bibfnamefont {J.}~\bibnamefont {Stenhammar}},
  \bibinfo {author} {\bibfnamefont {M.~E.}\ \bibnamefont {Cates}}, \bibinfo
  {author} {\bibfnamefont {Y.}~\bibnamefont {Kafri}},\ and\ \bibinfo {author}
  {\bibfnamefont {J.}~\bibnamefont {Tailleur}},\ }\bibfield  {journal}
  {\bibinfo  {journal} {Physical Review E}\ }\textbf {\bibinfo {volume} {97}},\
  \href {https://doi.org/10.1103/physreve.97.020602}
  {10.1103/physreve.97.020602} (\bibinfo {year} {2018})\BibitemShut {NoStop}%
\bibitem [{\citenamefont {Shankar}\ \emph {et~al.}(2018)\citenamefont
  {Shankar}, \citenamefont {Ramaswamy}, \citenamefont {Marchetti},\ and\
  \citenamefont {Bowick}}]{Shankar2018}%
  \BibitemOpen
  \bibfield  {author} {\bibinfo {author} {\bibfnamefont {S.}~\bibnamefont
  {Shankar}}, \bibinfo {author} {\bibfnamefont {S.}~\bibnamefont {Ramaswamy}},
  \bibinfo {author} {\bibfnamefont {M.~C.}\ \bibnamefont {Marchetti}},\ and\
  \bibinfo {author} {\bibfnamefont {M.~J.}\ \bibnamefont {Bowick}},\ }\bibfield
   {journal} {\bibinfo  {journal} {Physical Review Letters}\ }\textbf {\bibinfo
  {volume} {121}},\ \href {https://doi.org/10.1103/physrevlett.121.108002}
  {10.1103/physrevlett.121.108002} (\bibinfo {year} {2018})\BibitemShut
  {NoStop}%
\bibitem [{\citenamefont {J\"{u}licher}\ \emph {et~al.}(2018)\citenamefont
  {J\"{u}licher}, \citenamefont {Grill},\ and\ \citenamefont
  {Salbreux}}]{Juelicher2018}%
  \BibitemOpen
  \bibfield  {author} {\bibinfo {author} {\bibfnamefont {F.}~\bibnamefont
  {J\"{u}licher}}, \bibinfo {author} {\bibfnamefont {S.~W.}\ \bibnamefont
  {Grill}},\ and\ \bibinfo {author} {\bibfnamefont {G.}~\bibnamefont
  {Salbreux}},\ }\href {https://doi.org/10.1088/1361-6633/aab6bb} {\bibfield
  {journal} {\bibinfo  {journal} {Reports on Progress in Physics}\ }\textbf
  {\bibinfo {volume} {81}},\ \bibinfo {pages} {076601} (\bibinfo {year}
  {2018})}\BibitemShut {NoStop}%
\bibitem [{\citenamefont {Golestanian}(2019)}]{Golestanian2019}%
  \BibitemOpen
  \bibfield  {author} {\bibinfo {author} {\bibfnamefont {R.}~\bibnamefont
  {Golestanian}},\ }\href {https://doi.org/10.1103/PhysRevE.100.010601}
  {\bibfield  {journal} {\bibinfo  {journal} {Phys. Rev. E}\ }\textbf {\bibinfo
  {volume} {100}},\ \bibinfo {pages} {010601} (\bibinfo {year}
  {2019})}\BibitemShut {NoStop}%
\bibitem [{\citenamefont {Vicsek}\ \emph {et~al.}(1995)\citenamefont {Vicsek},
  \citenamefont {Czirók}, \citenamefont {Ben-Jacob}, \citenamefont {Cohen},\
  and\ \citenamefont {Shochet}}]{Vicsek1995}%
  \BibitemOpen
  \bibfield  {author} {\bibinfo {author} {\bibfnamefont {T.}~\bibnamefont
  {Vicsek}}, \bibinfo {author} {\bibfnamefont {A.}~\bibnamefont {Czirók}},
  \bibinfo {author} {\bibfnamefont {E.}~\bibnamefont {Ben-Jacob}}, \bibinfo
  {author} {\bibfnamefont {I.}~\bibnamefont {Cohen}},\ and\ \bibinfo {author}
  {\bibfnamefont {O.}~\bibnamefont {Shochet}},\ }\href
  {https://doi.org/10.1103/physrevlett.75.1226} {\bibfield  {journal} {\bibinfo
   {journal} {Physical Review Letters}\ }\textbf {\bibinfo {volume} {75}},\
  \bibinfo {pages} {1226–1229} (\bibinfo {year} {1995})}\BibitemShut
  {NoStop}%
\bibitem [{\citenamefont {Grégoire}\ and\ \citenamefont
  {Chaté}(2004)}]{Gregoire2004}%
  \BibitemOpen
  \bibfield  {author} {\bibinfo {author} {\bibfnamefont {G.}~\bibnamefont
  {Grégoire}}\ and\ \bibinfo {author} {\bibfnamefont {H.}~\bibnamefont
  {Chaté}},\ }\bibfield  {journal} {\bibinfo  {journal} {Physical Review
  Letters}\ }\textbf {\bibinfo {volume} {92}},\ \href
  {https://doi.org/10.1103/physrevlett.92.025702}
  {10.1103/physrevlett.92.025702} (\bibinfo {year} {2004})\BibitemShut
  {NoStop}%
\bibitem [{\citenamefont {Toner}\ and\ \citenamefont {Tu}(1995)}]{Toner1995}%
  \BibitemOpen
  \bibfield  {author} {\bibinfo {author} {\bibfnamefont {J.}~\bibnamefont
  {Toner}}\ and\ \bibinfo {author} {\bibfnamefont {Y.}~\bibnamefont {Tu}},\
  }\href {https://doi.org/10.1103/physrevlett.75.4326} {\bibfield  {journal}
  {\bibinfo  {journal} {Physical Review Letters}\ }\textbf {\bibinfo {volume}
  {75}},\ \bibinfo {pages} {4326–4329} (\bibinfo {year} {1995})}\BibitemShut
  {NoStop}%
\bibitem [{\citenamefont {Theurkauff}\ \emph {et~al.}(2012)\citenamefont
  {Theurkauff}, \citenamefont {Cottin-Bizonne}, \citenamefont {Palacci},
  \citenamefont {Ybert},\ and\ \citenamefont {Bocquet}}]{Theurkauff2012}%
  \BibitemOpen
  \bibfield  {author} {\bibinfo {author} {\bibfnamefont {I.}~\bibnamefont
  {Theurkauff}}, \bibinfo {author} {\bibfnamefont {C.}~\bibnamefont
  {Cottin-Bizonne}}, \bibinfo {author} {\bibfnamefont {J.}~\bibnamefont
  {Palacci}}, \bibinfo {author} {\bibfnamefont {C.}~\bibnamefont {Ybert}},\
  and\ \bibinfo {author} {\bibfnamefont {L.}~\bibnamefont {Bocquet}},\
  }\bibfield  {journal} {\bibinfo  {journal} {Physical Review Letters}\
  }\textbf {\bibinfo {volume} {108}},\ \href
  {https://doi.org/10.1103/physrevlett.108.268303}
  {10.1103/physrevlett.108.268303} (\bibinfo {year} {2012})\BibitemShut
  {NoStop}%
\bibitem [{\citenamefont {Buttinoni}\ \emph {et~al.}(2013)\citenamefont
  {Buttinoni}, \citenamefont {Bialké}, \citenamefont {K\"{u}mmel},
  \citenamefont {L\"{o}wen}, \citenamefont {Bechinger},\ and\ \citenamefont
  {Speck}}]{Buttinoni2013}%
  \BibitemOpen
  \bibfield  {author} {\bibinfo {author} {\bibfnamefont {I.}~\bibnamefont
  {Buttinoni}}, \bibinfo {author} {\bibfnamefont {J.}~\bibnamefont {Bialké}},
  \bibinfo {author} {\bibfnamefont {F.}~\bibnamefont {K\"{u}mmel}}, \bibinfo
  {author} {\bibfnamefont {H.}~\bibnamefont {L\"{o}wen}}, \bibinfo {author}
  {\bibfnamefont {C.}~\bibnamefont {Bechinger}},\ and\ \bibinfo {author}
  {\bibfnamefont {T.}~\bibnamefont {Speck}},\ }\bibfield  {journal} {\bibinfo
  {journal} {Physical Review Letters}\ }\textbf {\bibinfo {volume} {110}},\
  \href {https://doi.org/10.1103/physrevlett.110.238301}
  {10.1103/physrevlett.110.238301} (\bibinfo {year} {2013})\BibitemShut
  {NoStop}%
\bibitem [{\citenamefont {Hallatschek}\ \emph {et~al.}(2023)\citenamefont
  {Hallatschek}, \citenamefont {Datta}, \citenamefont {Drescher}, \citenamefont
  {Dunkel}, \citenamefont {Elgeti}, \citenamefont {Waclaw},\ and\ \citenamefont
  {Wingreen}}]{hallatschek_proliferating_2023}%
  \BibitemOpen
  \bibfield  {author} {\bibinfo {author} {\bibfnamefont {O.}~\bibnamefont
  {Hallatschek}}, \bibinfo {author} {\bibfnamefont {S.~S.}\ \bibnamefont
  {Datta}}, \bibinfo {author} {\bibfnamefont {K.}~\bibnamefont {Drescher}},
  \bibinfo {author} {\bibfnamefont {J.}~\bibnamefont {Dunkel}}, \bibinfo
  {author} {\bibfnamefont {J.}~\bibnamefont {Elgeti}}, \bibinfo {author}
  {\bibfnamefont {B.}~\bibnamefont {Waclaw}},\ and\ \bibinfo {author}
  {\bibfnamefont {N.~S.}\ \bibnamefont {Wingreen}},\ }\href
  {https://doi.org/10.1038/s42254-023-00593-0} {\bibfield  {journal} {\bibinfo
  {journal} {Nature Reviews Physics}\ ,\ \bibinfo {pages} {1}} (\bibinfo {year}
  {2023})},\ \bibinfo {note} {publisher: Nature Publishing Group}\BibitemShut
  {NoStop}%
\bibitem [{\citenamefont {Ranft}\ \emph {et~al.}(2010)\citenamefont {Ranft},
  \citenamefont {Basan}, \citenamefont {Elgeti}, \citenamefont {Joanny},
  \citenamefont {Prost},\ and\ \citenamefont {J\"{u}licher}}]{Ranft2010}%
  \BibitemOpen
  \bibfield  {author} {\bibinfo {author} {\bibfnamefont {J.}~\bibnamefont
  {Ranft}}, \bibinfo {author} {\bibfnamefont {M.}~\bibnamefont {Basan}},
  \bibinfo {author} {\bibfnamefont {J.}~\bibnamefont {Elgeti}}, \bibinfo
  {author} {\bibfnamefont {J.-F.}\ \bibnamefont {Joanny}}, \bibinfo {author}
  {\bibfnamefont {J.}~\bibnamefont {Prost}},\ and\ \bibinfo {author}
  {\bibfnamefont {F.}~\bibnamefont {J\"{u}licher}},\ }\href
  {https://doi.org/10.1073/pnas.1011086107} {\bibfield  {journal} {\bibinfo
  {journal} {Proceedings of the National Academy of Sciences}\ }\textbf
  {\bibinfo {volume} {107}},\ \bibinfo {pages} {20863–20868} (\bibinfo {year}
  {2010})}\BibitemShut {NoStop}%
\bibitem [{\citenamefont {Farrell}\ \emph {et~al.}(2013)\citenamefont
  {Farrell}, \citenamefont {Hallatschek}, \citenamefont {Marenduzzo},\ and\
  \citenamefont {Waclaw}}]{farrell_mechanically_2013}%
  \BibitemOpen
  \bibfield  {author} {\bibinfo {author} {\bibfnamefont {F.~D.~C.}\
  \bibnamefont {Farrell}}, \bibinfo {author} {\bibfnamefont {O.}~\bibnamefont
  {Hallatschek}}, \bibinfo {author} {\bibfnamefont {D.}~\bibnamefont
  {Marenduzzo}},\ and\ \bibinfo {author} {\bibfnamefont {B.}~\bibnamefont
  {Waclaw}},\ }\bibfield  {journal} {\bibinfo  {journal} {Physical Review
  Letters}\ }\textbf {\bibinfo {volume} {111}},\ \href
  {https://doi.org/10.1103/PhysRevLett.111.168101}
  {10.1103/PhysRevLett.111.168101} (\bibinfo {year} {2013})\BibitemShut
  {NoStop}%
\bibitem [{\citenamefont {Grant}\ \emph {et~al.}(2014)\citenamefont {Grant},
  \citenamefont {Wacław}, \citenamefont {Allen},\ and\ \citenamefont
  {Cicuta}}]{Grant2014}%
  \BibitemOpen
  \bibfield  {author} {\bibinfo {author} {\bibfnamefont {M.~A.~A.}\
  \bibnamefont {Grant}}, \bibinfo {author} {\bibfnamefont {B.}~\bibnamefont
  {Wacław}}, \bibinfo {author} {\bibfnamefont {R.~J.}\ \bibnamefont {Allen}},\
  and\ \bibinfo {author} {\bibfnamefont {P.}~\bibnamefont {Cicuta}},\ }\href
  {https://doi.org/10.1098/rsif.2014.0400} {\bibfield  {journal} {\bibinfo
  {journal} {Journal of The Royal Society Interface}\ }\textbf {\bibinfo
  {volume} {11}},\ \bibinfo {pages} {20140400} (\bibinfo {year}
  {2014})}\BibitemShut {NoStop}%
\bibitem [{\citenamefont {Dell’Arciprete}\ \emph {et~al.}(2018)\citenamefont
  {Dell’Arciprete}, \citenamefont {Blow}, \citenamefont {Brown},
  \citenamefont {Farrell}, \citenamefont {Lintuvuori}, \citenamefont {McVey},
  \citenamefont {Marenduzzo},\ and\ \citenamefont {Poon}}]{DellArciprete2018}%
  \BibitemOpen
  \bibfield  {author} {\bibinfo {author} {\bibfnamefont {D.}~\bibnamefont
  {Dell’Arciprete}}, \bibinfo {author} {\bibfnamefont {M.~L.}\ \bibnamefont
  {Blow}}, \bibinfo {author} {\bibfnamefont {A.~T.}\ \bibnamefont {Brown}},
  \bibinfo {author} {\bibfnamefont {F.~D.~C.}\ \bibnamefont {Farrell}},
  \bibinfo {author} {\bibfnamefont {J.~S.}\ \bibnamefont {Lintuvuori}},
  \bibinfo {author} {\bibfnamefont {A.~F.}\ \bibnamefont {McVey}}, \bibinfo
  {author} {\bibfnamefont {D.}~\bibnamefont {Marenduzzo}},\ and\ \bibinfo
  {author} {\bibfnamefont {W.~C.~K.}\ \bibnamefont {Poon}},\ }\bibfield
  {journal} {\bibinfo  {journal} {Nature Communications}\ }\textbf {\bibinfo
  {volume} {9}},\ \href {https://doi.org/10.1038/s41467-018-06370-3}
  {10.1038/s41467-018-06370-3} (\bibinfo {year} {2018})\BibitemShut {NoStop}%
\bibitem [{\citenamefont {Yaman}\ \emph {et~al.}(2019)\citenamefont {Yaman},
  \citenamefont {Demir}, \citenamefont {Vetter},\ and\ \citenamefont
  {Kocabas}}]{Yaman2019}%
  \BibitemOpen
  \bibfield  {author} {\bibinfo {author} {\bibfnamefont {Y.~I.}\ \bibnamefont
  {Yaman}}, \bibinfo {author} {\bibfnamefont {E.}~\bibnamefont {Demir}},
  \bibinfo {author} {\bibfnamefont {R.}~\bibnamefont {Vetter}},\ and\ \bibinfo
  {author} {\bibfnamefont {A.}~\bibnamefont {Kocabas}},\ }\bibfield  {journal}
  {\bibinfo  {journal} {Nature Communications}\ }\textbf {\bibinfo {volume}
  {10}},\ \href {https://doi.org/10.1038/s41467-019-10311-z}
  {10.1038/s41467-019-10311-z} (\bibinfo {year} {2019})\BibitemShut {NoStop}%
\bibitem [{\citenamefont {Zhang}\ \emph {et~al.}(2021)\citenamefont {Zhang},
  \citenamefont {Li}, \citenamefont {Nijjer}, \citenamefont {Lu}, \citenamefont
  {Kothari}, \citenamefont {Alert}, \citenamefont {Cohen},\ and\ \citenamefont
  {Yan}}]{Zhang2021}%
  \BibitemOpen
  \bibfield  {author} {\bibinfo {author} {\bibfnamefont {Q.}~\bibnamefont
  {Zhang}}, \bibinfo {author} {\bibfnamefont {J.}~\bibnamefont {Li}}, \bibinfo
  {author} {\bibfnamefont {J.}~\bibnamefont {Nijjer}}, \bibinfo {author}
  {\bibfnamefont {H.}~\bibnamefont {Lu}}, \bibinfo {author} {\bibfnamefont
  {M.}~\bibnamefont {Kothari}}, \bibinfo {author} {\bibfnamefont
  {R.}~\bibnamefont {Alert}}, \bibinfo {author} {\bibfnamefont
  {T.}~\bibnamefont {Cohen}},\ and\ \bibinfo {author} {\bibfnamefont
  {J.}~\bibnamefont {Yan}},\ }\bibfield  {journal} {\bibinfo  {journal}
  {Proceedings of the National Academy of Sciences}\ }\textbf {\bibinfo
  {volume} {118}},\ \href {https://doi.org/10.1073/pnas.2107107118}
  {10.1073/pnas.2107107118} (\bibinfo {year} {2021})\BibitemShut {NoStop}%
\bibitem [{\citenamefont {Pollack}\ \emph {et~al.}(2022)\citenamefont
  {Pollack}, \citenamefont {Bittihn},\ and\ \citenamefont
  {Golestanian}}]{Pollack2022}%
  \BibitemOpen
  \bibfield  {author} {\bibinfo {author} {\bibfnamefont {Y.~G.}\ \bibnamefont
  {Pollack}}, \bibinfo {author} {\bibfnamefont {P.}~\bibnamefont {Bittihn}},\
  and\ \bibinfo {author} {\bibfnamefont {R.}~\bibnamefont {Golestanian}},\
  }\href {https://doi.org/10.1088/1367-2630/ac788e} {\bibfield  {journal}
  {\bibinfo  {journal} {New Journal of Physics}\ }\textbf {\bibinfo {volume}
  {24}},\ \bibinfo {pages} {073003} (\bibinfo {year} {2022})}\BibitemShut
  {NoStop}%
\bibitem [{\citenamefont {Isensee}\ \emph {et~al.}(2022)\citenamefont
  {Isensee}, \citenamefont {Hupe}, \citenamefont {Golestanian},\ and\
  \citenamefont {Bittihn}}]{Isensee2022}%
  \BibitemOpen
  \bibfield  {author} {\bibinfo {author} {\bibfnamefont {J.}~\bibnamefont
  {Isensee}}, \bibinfo {author} {\bibfnamefont {L.}~\bibnamefont {Hupe}},
  \bibinfo {author} {\bibfnamefont {R.}~\bibnamefont {Golestanian}},\ and\
  \bibinfo {author} {\bibfnamefont {P.}~\bibnamefont {Bittihn}},\ }\bibfield
  {journal} {\bibinfo  {journal} {Journal of The Royal Society Interface}\
  }\textbf {\bibinfo {volume} {19}},\ \href
  {https://doi.org/10.1098/rsif.2022.0512} {10.1098/rsif.2022.0512} (\bibinfo
  {year} {2022})\BibitemShut {NoStop}%
\bibitem [{\citenamefont {Steinberg}\ \emph {et~al.}(2020)\citenamefont
  {Steinberg}, \citenamefont {Keren-Paz}, \citenamefont {Hou}, \citenamefont
  {Doron}, \citenamefont {Yanuka-Golub}, \citenamefont {Olender}, \citenamefont
  {Hadar}, \citenamefont {Rosenberg}, \citenamefont {Jain}, \citenamefont
  {Cámara-Almirón}, \citenamefont {Romero}, \citenamefont {van Teeffelen},\
  and\ \citenamefont {Kolodkin-Gal}}]{Steinberg2020MotileSubpopulation}%
  \BibitemOpen
  \bibfield  {author} {\bibinfo {author} {\bibfnamefont {N.}~\bibnamefont
  {Steinberg}}, \bibinfo {author} {\bibfnamefont {A.}~\bibnamefont
  {Keren-Paz}}, \bibinfo {author} {\bibfnamefont {Q.}~\bibnamefont {Hou}},
  \bibinfo {author} {\bibfnamefont {S.}~\bibnamefont {Doron}}, \bibinfo
  {author} {\bibfnamefont {K.}~\bibnamefont {Yanuka-Golub}}, \bibinfo {author}
  {\bibfnamefont {T.}~\bibnamefont {Olender}}, \bibinfo {author} {\bibfnamefont
  {R.}~\bibnamefont {Hadar}}, \bibinfo {author} {\bibfnamefont
  {G.}~\bibnamefont {Rosenberg}}, \bibinfo {author} {\bibfnamefont
  {R.}~\bibnamefont {Jain}}, \bibinfo {author} {\bibfnamefont {J.}~\bibnamefont
  {Cámara-Almirón}}, \bibinfo {author} {\bibfnamefont {D.}~\bibnamefont
  {Romero}}, \bibinfo {author} {\bibfnamefont {S.}~\bibnamefont {van
  Teeffelen}},\ and\ \bibinfo {author} {\bibfnamefont {I.}~\bibnamefont
  {Kolodkin-Gal}},\ }\href {https://doi.org/10.1126/scisignal.aaw8905}
  {\bibfield  {journal} {\bibinfo  {journal} {Science Signaling}\ }\textbf
  {\bibinfo {volume} {13}},\ \bibinfo {pages} {eaaw8905} (\bibinfo {year}
  {2020})},\ \Eprint
  {https://arxiv.org/abs/https://www.science.org/doi/pdf/10.1126/scisignal.aaw8905}
  {https://www.science.org/doi/pdf/10.1126/scisignal.aaw8905} \BibitemShut
  {NoStop}%
\bibitem [{\citenamefont {Ravel}\ \emph {et~al.}(2022)\citenamefont {Ravel},
  \citenamefont {Bergmann}, \citenamefont {Trubuil}, \citenamefont {Deschamps},
  \citenamefont {Briandet},\ and\ \citenamefont
  {Labarthe}}]{Ravel2022swimming}%
  \BibitemOpen
  \bibfield  {author} {\bibinfo {author} {\bibfnamefont {G.}~\bibnamefont
  {Ravel}}, \bibinfo {author} {\bibfnamefont {M.}~\bibnamefont {Bergmann}},
  \bibinfo {author} {\bibfnamefont {A.}~\bibnamefont {Trubuil}}, \bibinfo
  {author} {\bibfnamefont {J.}~\bibnamefont {Deschamps}}, \bibinfo {author}
  {\bibfnamefont {R.}~\bibnamefont {Briandet}},\ and\ \bibinfo {author}
  {\bibfnamefont {S.}~\bibnamefont {Labarthe}},\ }\href
  {https://doi.org/10.7554/eLife.76513} {\bibfield  {journal} {\bibinfo
  {journal} {eLife}\ }\textbf {\bibinfo {volume} {11}},\ \bibinfo {pages}
  {e76513} (\bibinfo {year} {2022})}\BibitemShut {NoStop}%
\bibitem [{\citenamefont {Thiery}(2002)}]{thiery_epithelialmesenchymal_2002}%
  \BibitemOpen
  \bibfield  {author} {\bibinfo {author} {\bibfnamefont {J.~P.}\ \bibnamefont
  {Thiery}},\ }\href {https://doi.org/10.1038/nrc822} {\bibfield  {journal}
  {\bibinfo  {journal} {Nature Reviews Cancer}\ }\textbf {\bibinfo {volume}
  {2}},\ \bibinfo {pages} {442} (\bibinfo {year} {2002})}\BibitemShut {NoStop}%
\bibitem [{\citenamefont {Stenhammar}\ \emph {et~al.}(2015)\citenamefont
  {Stenhammar}, \citenamefont {Wittkowski}, \citenamefont {Marenduzzo},\ and\
  \citenamefont {Cates}}]{Stenhammar_activepassive_2015}%
  \BibitemOpen
  \bibfield  {author} {\bibinfo {author} {\bibfnamefont {J.}~\bibnamefont
  {Stenhammar}}, \bibinfo {author} {\bibfnamefont {R.}~\bibnamefont
  {Wittkowski}}, \bibinfo {author} {\bibfnamefont {D.}~\bibnamefont
  {Marenduzzo}},\ and\ \bibinfo {author} {\bibfnamefont {M.~E.}\ \bibnamefont
  {Cates}},\ }\href {https://doi.org/10.1103/PhysRevLett.114.018301} {\bibfield
   {journal} {\bibinfo  {journal} {Phys. Rev. Lett.}\ }\textbf {\bibinfo
  {volume} {114}},\ \bibinfo {pages} {018301} (\bibinfo {year}
  {2015})}\BibitemShut {NoStop}%
\bibitem [{\citenamefont {Wysocki}\ \emph {et~al.}(2016)\citenamefont
  {Wysocki}, \citenamefont {Winkler},\ and\ \citenamefont
  {Gompper}}]{Wysocki2016}%
  \BibitemOpen
  \bibfield  {author} {\bibinfo {author} {\bibfnamefont {A.}~\bibnamefont
  {Wysocki}}, \bibinfo {author} {\bibfnamefont {R.~G.}\ \bibnamefont
  {Winkler}},\ and\ \bibinfo {author} {\bibfnamefont {G.}~\bibnamefont
  {Gompper}},\ }\href {https://doi.org/10.1088/1367-2630/aa529d} {\bibfield
  {journal} {\bibinfo  {journal} {New Journal of Physics}\ }\textbf {\bibinfo
  {volume} {18}},\ \bibinfo {pages} {123030} (\bibinfo {year}
  {2016})}\BibitemShut {NoStop}%
\bibitem [{\citenamefont {Dolai}\ \emph {et~al.}(2018)\citenamefont {Dolai},
  \citenamefont {Simha},\ and\ \citenamefont
  {Mishra}}]{Dolai_activepassive_2018}%
  \BibitemOpen
  \bibfield  {author} {\bibinfo {author} {\bibfnamefont {P.}~\bibnamefont
  {Dolai}}, \bibinfo {author} {\bibfnamefont {A.}~\bibnamefont {Simha}},\ and\
  \bibinfo {author} {\bibfnamefont {S.}~\bibnamefont {Mishra}},\ }\href
  {https://doi.org/10.1039/C8SM00222C} {\bibfield  {journal} {\bibinfo
  {journal} {Soft Matter}\ }\textbf {\bibinfo {volume} {14}},\ \bibinfo {pages}
  {6137} (\bibinfo {year} {2018})}\BibitemShut {NoStop}%
\bibitem [{\citenamefont {McCandlish}\ \emph {et~al.}(2012)\citenamefont
  {McCandlish}, \citenamefont {Baskaran},\ and\ \citenamefont
  {Hagan}}]{McCandlish2012}%
  \BibitemOpen
  \bibfield  {author} {\bibinfo {author} {\bibfnamefont {S.~R.}\ \bibnamefont
  {McCandlish}}, \bibinfo {author} {\bibfnamefont {A.}~\bibnamefont
  {Baskaran}},\ and\ \bibinfo {author} {\bibfnamefont {M.~F.}\ \bibnamefont
  {Hagan}},\ }\href {https://doi.org/10.1039/c2sm06960a} {\bibfield  {journal}
  {\bibinfo  {journal} {Soft Matter}\ }\textbf {\bibinfo {volume} {8}},\
  \bibinfo {pages} {2527} (\bibinfo {year} {2012})}\BibitemShut {NoStop}%
\bibitem [{\citenamefont {Saha}\ \emph {et~al.}(2020)\citenamefont {Saha},
  \citenamefont {Agudo-Canalejo},\ and\ \citenamefont
  {Golestanian}}]{Saha2020}%
  \BibitemOpen
  \bibfield  {author} {\bibinfo {author} {\bibfnamefont {S.}~\bibnamefont
  {Saha}}, \bibinfo {author} {\bibfnamefont {J.}~\bibnamefont
  {Agudo-Canalejo}},\ and\ \bibinfo {author} {\bibfnamefont {R.}~\bibnamefont
  {Golestanian}},\ }\bibfield  {journal} {\bibinfo  {journal} {Physical Review
  X}\ }\textbf {\bibinfo {volume} {10}},\ \href
  {https://doi.org/10.1103/physrevx.10.041009} {10.1103/physrevx.10.041009}
  (\bibinfo {year} {2020})\BibitemShut {NoStop}%
\bibitem [{\citenamefont {Ouazan-Reboul}\ \emph {et~al.}(2023)\citenamefont
  {Ouazan-Reboul}, \citenamefont {Agudo-Canalejo},\ and\ \citenamefont
  {Golestanian}}]{OuazanReboul_metaboliccycles_2023}%
  \BibitemOpen
  \bibfield  {author} {\bibinfo {author} {\bibfnamefont {V.}~\bibnamefont
  {Ouazan-Reboul}}, \bibinfo {author} {\bibfnamefont {J.}~\bibnamefont
  {Agudo-Canalejo}},\ and\ \bibinfo {author} {\bibfnamefont {R.}~\bibnamefont
  {Golestanian}},\ }\href {https://doi.org/10.1038/s41467-023-40241-w}
  {\bibfield  {journal} {\bibinfo  {journal} {Nature Communications}\ }\textbf
  {\bibinfo {volume} {14}},\ \bibinfo {pages} {4496} (\bibinfo {year}
  {2023})}\BibitemShut {NoStop}%
\bibitem [{\citenamefont {Dinelli}\ \emph {et~al.}(2023)\citenamefont
  {Dinelli}, \citenamefont {O’Byrne}, \citenamefont {Curatolo}, \citenamefont
  {Zhao}, \citenamefont {Sollich},\ and\ \citenamefont
  {Tailleur}}]{Dinelli_nonreciprocity_2023}%
  \BibitemOpen
  \bibfield  {author} {\bibinfo {author} {\bibfnamefont {A.}~\bibnamefont
  {Dinelli}}, \bibinfo {author} {\bibfnamefont {J.}~\bibnamefont {O’Byrne}},
  \bibinfo {author} {\bibfnamefont {A.}~\bibnamefont {Curatolo}}, \bibinfo
  {author} {\bibfnamefont {Y.}~\bibnamefont {Zhao}}, \bibinfo {author}
  {\bibfnamefont {P.}~\bibnamefont {Sollich}},\ and\ \bibinfo {author}
  {\bibfnamefont {J.}~\bibnamefont {Tailleur}},\ }\href
  {https://doi.org/10.1038/s41467-023-42713-5} {\bibfield  {journal} {\bibinfo
  {journal} {Nature Communications}\ }\textbf {\bibinfo {volume} {14}},\
  \bibinfo {pages} {7035} (\bibinfo {year} {2023})}\BibitemShut {NoStop}%
\bibitem [{\citenamefont {Ghosh}\ \emph {et~al.}(2015)\citenamefont {Ghosh},
  \citenamefont {Mondal}, \citenamefont {Ben-Jacob},\ and\ \citenamefont
  {Levine}}]{ghosh_mechanically-driven_2015}%
  \BibitemOpen
  \bibfield  {author} {\bibinfo {author} {\bibfnamefont {P.}~\bibnamefont
  {Ghosh}}, \bibinfo {author} {\bibfnamefont {J.}~\bibnamefont {Mondal}},
  \bibinfo {author} {\bibfnamefont {E.}~\bibnamefont {Ben-Jacob}},\ and\
  \bibinfo {author} {\bibfnamefont {H.}~\bibnamefont {Levine}},\ }\href
  {http://www.pnas.org/content/early/2015/04/08/1504948112.short} {\bibfield
  {journal} {\bibinfo  {journal} {Proceedings of the National Academy of
  Sciences}\ }\textbf {\bibinfo {volume} {112}},\ \bibinfo {pages} {201504948}
  (\bibinfo {year} {2015})}\BibitemShut {NoStop}%
\bibitem [{\citenamefont {Kolb}\ and\ \citenamefont
  {Klotsa}(2020)}]{kolb_active_2020}%
  \BibitemOpen
  \bibfield  {author} {\bibinfo {author} {\bibfnamefont {T.}~\bibnamefont
  {Kolb}}\ and\ \bibinfo {author} {\bibfnamefont {D.}~\bibnamefont {Klotsa}},\
  }\href {https://doi.org/10.1039/C9SM01799B} {\bibfield  {journal} {\bibinfo
  {journal} {Soft Matter}\ }\textbf {\bibinfo {volume} {16}},\ \bibinfo {pages}
  {1967} (\bibinfo {year} {2020})},\ \bibinfo {note} {publisher: The Royal
  Society of Chemistry}\BibitemShut {NoStop}%
\bibitem [{\citenamefont {Braat}\ \emph {et~al.}(2024)\citenamefont {Braat},
  \citenamefont {Storm},\ and\ \citenamefont {Janssen}}]{Braat2024}%
  \BibitemOpen
  \bibfield  {author} {\bibinfo {author} {\bibfnamefont {Q.~J.~S.}\
  \bibnamefont {Braat}}, \bibinfo {author} {\bibfnamefont {C.}~\bibnamefont
  {Storm}},\ and\ \bibinfo {author} {\bibfnamefont {L.~M.~C.}\ \bibnamefont
  {Janssen}},\ }\bibfield  {journal} {\bibinfo  {journal} {Physical Review E}\
  }\textbf {\bibinfo {volume} {110}},\ \href
  {https://doi.org/10.1103/physreve.110.064401} {10.1103/physreve.110.064401}
  (\bibinfo {year} {2024})\BibitemShut {NoStop}%
\bibitem [{\citenamefont {Hupe*}\ \emph {et~al.}(2024)\citenamefont {Hupe*},
  \citenamefont {Pollack*}, \citenamefont {Isensee}, \citenamefont {Amiri},
  \citenamefont {Golestanian},\ and\ \citenamefont
  {Bittihn}}]{hupe_minimal_2024}%
  \BibitemOpen
  \bibfield  {author} {\bibinfo {author} {\bibfnamefont {L.}~\bibnamefont
  {Hupe*}}, \bibinfo {author} {\bibfnamefont {Y.~G.}\ \bibnamefont {Pollack*}},
  \bibinfo {author} {\bibfnamefont {J.}~\bibnamefont {Isensee}}, \bibinfo
  {author} {\bibfnamefont {A.}~\bibnamefont {Amiri}}, \bibinfo {author}
  {\bibfnamefont {R.}~\bibnamefont {Golestanian}},\ and\ \bibinfo {author}
  {\bibfnamefont {P.}~\bibnamefont {Bittihn}},\ }\href
  {https://doi.org/10.48550/arXiv.2409.01959} {\bibinfo {title} {A minimal
  model of smoothly dividing disk-shaped cells}} (\bibinfo {year} {2024}),\
  \Eprint {https://arxiv.org/abs/2409.01959} {arXiv:2409.01959 [cond-mat]}
  \BibitemShut {NoStop}%
\bibitem [{Sup()}]{SuppMat}%
  \BibitemOpen
  \href@noop {} {\bibinfo {title} {See supplemental material for details on the
  numerical model, the clustering analysis, and the fitting procedure for the
  effective potential, as well as supporting evidence for the coarsening
  dynamics and bath dynamics near larger objects.}}\BibitemShut {Stop}%
\bibitem [{Note1()}]{Note1}%
  \BibitemOpen
  \bibinfo {note} {To keep the analysis consistent with subsequent velocity
  calculations, we take the spatial location to be defined by the corresponding
  particles' center positions}\BibitemShut {NoStop}%
\bibitem [{\citenamefont {Howse}\ \emph {et~al.}(2007)\citenamefont {Howse},
  \citenamefont {Jones}, \citenamefont {Ryan}, \citenamefont {Gough},
  \citenamefont {Vafabakhsh},\ and\ \citenamefont {Golestanian}}]{Howse2007}%
  \BibitemOpen
  \bibfield  {author} {\bibinfo {author} {\bibfnamefont {J.~R.}\ \bibnamefont
  {Howse}}, \bibinfo {author} {\bibfnamefont {R.~A.~L.}\ \bibnamefont {Jones}},
  \bibinfo {author} {\bibfnamefont {A.~J.}\ \bibnamefont {Ryan}}, \bibinfo
  {author} {\bibfnamefont {T.}~\bibnamefont {Gough}}, \bibinfo {author}
  {\bibfnamefont {R.}~\bibnamefont {Vafabakhsh}},\ and\ \bibinfo {author}
  {\bibfnamefont {R.}~\bibnamefont {Golestanian}},\ }\href
  {https://doi.org/10.1103/PhysRevLett.99.048102} {\bibfield  {journal}
  {\bibinfo  {journal} {Phys. Rev. Lett.}\ }\textbf {\bibinfo {volume} {99}},\
  \bibinfo {pages} {048102} (\bibinfo {year} {2007})}\BibitemShut {NoStop}%
\bibitem [{\citenamefont {Bechinger}\ \emph {et~al.}(2016)\citenamefont
  {Bechinger}, \citenamefont {Di~Leonardo}, \citenamefont {L{\"o}wen},
  \citenamefont {Reichhardt}, \citenamefont {Volpe},\ and\ \citenamefont
  {Volpe}}]{bechinger_active_2016}%
  \BibitemOpen
  \bibfield  {author} {\bibinfo {author} {\bibfnamefont {C.}~\bibnamefont
  {Bechinger}}, \bibinfo {author} {\bibfnamefont {R.}~\bibnamefont
  {Di~Leonardo}}, \bibinfo {author} {\bibfnamefont {H.}~\bibnamefont
  {L{\"o}wen}}, \bibinfo {author} {\bibfnamefont {C.}~\bibnamefont
  {Reichhardt}}, \bibinfo {author} {\bibfnamefont {G.}~\bibnamefont {Volpe}},\
  and\ \bibinfo {author} {\bibfnamefont {G.}~\bibnamefont {Volpe}},\ }\href
  {https://doi.org/10.1103/RevModPhys.88.045006} {\bibfield  {journal}
  {\bibinfo  {journal} {Reviews of Modern Physics}\ }\textbf {\bibinfo {volume}
  {88}},\ \bibinfo {pages} {045006} (\bibinfo {year} {2016})}\BibitemShut
  {NoStop}%
\bibitem [{\citenamefont {Sunkel}\ \emph {et~al.}(2025)\citenamefont {Sunkel},
  \citenamefont {Hupe},\ and\ \citenamefont
  {Bittihn}}]{sunkel_motilityinduced_2025}%
  \BibitemOpen
  \bibfield  {author} {\bibinfo {author} {\bibfnamefont {T.}~\bibnamefont
  {Sunkel}}, \bibinfo {author} {\bibfnamefont {L.}~\bibnamefont {Hupe}},\ and\
  \bibinfo {author} {\bibfnamefont {P.}~\bibnamefont {Bittihn}},\ }\href
  {https://doi.org/10.1038/s42005-025-02090-5} {\bibfield  {journal} {\bibinfo
  {journal} {Communications Physics}\ }\textbf {\bibinfo {volume} {8}},\
  \bibinfo {pages} {1} (\bibinfo {year} {2025})}\BibitemShut {NoStop}%
\bibitem [{\citenamefont {Lish}\ \emph {et~al.}(2024)\citenamefont {Lish},
  \citenamefont {Hupe}, \citenamefont {Golestanian},\ and\ \citenamefont
  {Bittihn}}]{lish_isovolumetric_2024}%
  \BibitemOpen
  \bibfield  {author} {\bibinfo {author} {\bibfnamefont {S.~R.}\ \bibnamefont
  {Lish}}, \bibinfo {author} {\bibfnamefont {L.}~\bibnamefont {Hupe}}, \bibinfo
  {author} {\bibfnamefont {R.}~\bibnamefont {Golestanian}},\ and\ \bibinfo
  {author} {\bibfnamefont {P.}~\bibnamefont {Bittihn}},\ }\href
  {https://doi.org/10.48550/arXiv.2409.20481} {\bibinfo {title} {Isovolumetric
  dividing active matter}} (\bibinfo {year} {2024}),\ \Eprint
  {https://arxiv.org/abs/2409.20481} {arXiv:2409.20481 [cond-mat]} \BibitemShut
  {NoStop}%
\bibitem [{\citenamefont {Kardar}\ and\ \citenamefont
  {Golestanian}(1999)}]{Casimir1999}%
  \BibitemOpen
  \bibfield  {author} {\bibinfo {author} {\bibfnamefont {M.}~\bibnamefont
  {Kardar}}\ and\ \bibinfo {author} {\bibfnamefont {R.}~\bibnamefont
  {Golestanian}},\ }\href {https://doi.org/10.1103/RevModPhys.71.1233}
  {\bibfield  {journal} {\bibinfo  {journal} {Rev. Mod. Phys.}\ }\textbf
  {\bibinfo {volume} {71}},\ \bibinfo {pages} {1233} (\bibinfo {year}
  {1999})}\BibitemShut {NoStop}%
\bibitem [{\citenamefont {Mahault}\ and\ \citenamefont
  {Golestanian}(2020)}]{Mahault2020}%
  \BibitemOpen
  \bibfield  {author} {\bibinfo {author} {\bibfnamefont {B.}~\bibnamefont
  {Mahault}}\ and\ \bibinfo {author} {\bibfnamefont {R.}~\bibnamefont
  {Golestanian}},\ }\href {https://doi.org/10.1088/1367-2630/ab90d8} {\bibfield
   {journal} {\bibinfo  {journal} {New Journal of Physics}\ }\textbf {\bibinfo
  {volume} {22}},\ \bibinfo {pages} {063045} (\bibinfo {year}
  {2020})}\BibitemShut {NoStop}%
\bibitem [{\citenamefont {Waclaw}\ \emph {et~al.}(2009)\citenamefont {Waclaw},
  \citenamefont {Sopik}, \citenamefont {Janke},\ and\ \citenamefont
  {Meyer-Ortmanns}}]{waclaw_mass_2009}%
  \BibitemOpen
  \bibfield  {author} {\bibinfo {author} {\bibfnamefont {B.}~\bibnamefont
  {Waclaw}}, \bibinfo {author} {\bibfnamefont {J.}~\bibnamefont {Sopik}},
  \bibinfo {author} {\bibfnamefont {W.}~\bibnamefont {Janke}},\ and\ \bibinfo
  {author} {\bibfnamefont {H.}~\bibnamefont {Meyer-Ortmanns}},\ }\href
  {https://doi.org/10.1088/1742-5468/2009/10/P10021} {\bibfield  {journal}
  {\bibinfo  {journal} {Journal of Statistical Mechanics: Theory and
  Experiment}\ }\textbf {\bibinfo {volume} {2009}},\ \bibinfo {pages} {P10021}
  (\bibinfo {year} {2009})}\BibitemShut {NoStop}%
\bibitem [{\citenamefont {Waclaw}\ and\ \citenamefont
  {Evans}(2012)}]{waclaw_explosive_2012}%
  \BibitemOpen
  \bibfield  {author} {\bibinfo {author} {\bibfnamefont {B.}~\bibnamefont
  {Waclaw}}\ and\ \bibinfo {author} {\bibfnamefont {M.~R.}\ \bibnamefont
  {Evans}},\ }\href {https://doi.org/10.1103/PhysRevLett.108.070601} {\bibfield
   {journal} {\bibinfo  {journal} {Physical Review Letters}\ }\textbf {\bibinfo
  {volume} {108}},\ \bibinfo {pages} {070601} (\bibinfo {year}
  {2012})}\BibitemShut {NoStop}%
\bibitem [{\citenamefont {Bezanson}\ \emph {et~al.}(2017)\citenamefont
  {Bezanson}, \citenamefont {Edelman}, \citenamefont {Karpinski},\ and\
  \citenamefont {Shah}}]{bezanson_julia_2017}%
  \BibitemOpen
  \bibfield  {author} {\bibinfo {author} {\bibfnamefont {J.}~\bibnamefont
  {Bezanson}}, \bibinfo {author} {\bibfnamefont {A.}~\bibnamefont {Edelman}},
  \bibinfo {author} {\bibfnamefont {S.}~\bibnamefont {Karpinski}},\ and\
  \bibinfo {author} {\bibfnamefont {V.}~\bibnamefont {Shah}},\ }\href
  {https://doi.org/10.1137/141000671} {\bibfield  {journal} {\bibinfo
  {journal} {SIAM Rev.}\ }\textbf {\bibinfo {volume} {59}},\ \bibinfo {pages}
  {65} (\bibinfo {year} {2017})}\BibitemShut {NoStop}%
\bibitem [{\citenamefont {Hupe}\ \emph {et~al.}(2022)\citenamefont {Hupe},
  \citenamefont {Isensee},\ and\ \citenamefont {Bittihn}}]{hupe_inparts_2022}%
  \BibitemOpen
  \bibfield  {author} {\bibinfo {author} {\bibfnamefont {L.}~\bibnamefont
  {Hupe}}, \bibinfo {author} {\bibfnamefont {J.}~\bibnamefont {Isensee}},\ and\
  \bibinfo {author} {\bibfnamefont {P.}~\bibnamefont {Bittihn}},\ }\href
  {http://hdl.handle.net/21.11101/0000-0007-FE12-7} {\bibinfo {title}
  {{InPartS}: {Interacting} {Particle} {Simulations} in {Julia},
  \href{https://www.inparts.org}{InPartS.org}}} (\bibinfo {year}
  {2022})\BibitemShut {NoStop}%
\bibitem [{\citenamefont {Danisch}\ and\ \citenamefont
  {Krumbiegel}(2021)}]{danisch_makie_2021}%
  \BibitemOpen
  \bibfield  {author} {\bibinfo {author} {\bibfnamefont {S.}~\bibnamefont
  {Danisch}}\ and\ \bibinfo {author} {\bibfnamefont {J.}~\bibnamefont
  {Krumbiegel}},\ }\href {https://doi.org/10.21105/joss.03349} {\bibfield
  {journal} {\bibinfo  {journal} {Journal of Open Source Software}\ }\textbf
  {\bibinfo {volume} {6}},\ \bibinfo {pages} {3349} (\bibinfo {year}
  {2021})}\BibitemShut {NoStop}%
\end{thebibliography}
\end{document}